\documentclass[10pt]{article}
\pdfoutput=1
\usepackage{jhep-mod-coarse}
\usepackage{bm}
\usepackage{amssymb}% http://ctan.org/pkg/amssymb
\usepackage{hyperref}
\usepackage{color}
\def\d{{\mathrm{d}}}
\def\g{{\mathfrak{g}}}

\def\lint{\hbox{\Large $\displaystyle\int$}}   %needs \usepackage{amssymb} [large integral]
\def\hint{\hbox{\huge $\displaystyle\int$}}  %needs \usepackage{amssymb} [huge integral]
\makeatletter
\newcommand{\stkout}[1]{\ifmmode\text{\sout{\ensuremath{#1}}}\else\sout{#1}\fi}
\begin{document}
%------------------------------------------------------------------------------------------------------------------------------------------
%=================================================================
% Our local definitions
%=================================================================
\def\d{{\mathrm{d}}}
\def\g{{\mathfrak{g}}}
\def\g{{\mathfrak{g}}}
\def\dbar{{\mathchar'26\mkern-12mu \d}} 
\def\tr{{\mathrm{tr}}}
\def\Hilbert{{\mathcal{H}}}
\def\H{{\mathrm{H}}}
\def\R{{\mathrm{R}}}
\def\E{{\mathrm{E}}}
\def\ss{{\hbox{\textdollaroldstyle}}}
%=================================================================
\title{\leftline{Coarse graining Shannon and von Neumann entropies}}
%=================================================================
\author{\large Ana Alonso-Serrano $^{1}$ {\sf and}  Matt Visser $^{2}$}
%=================================================================
%=================================================================
\affiliation{%
$^{1}$ Institute of Theoretical Physics, Faculty of Mathematics and Physics, Charles University, \\
\null\qquad 1800 Prague, Czech Republic;  {\sf a.alonso.serrano@utf.mff.cuni.cz}\\
$^{2}$ School of Mathematics and Statistics, Victoria University of Wellington, \\
\null\qquad PO Box 600, Wellington 6140, 
New Zealand; {\sf matt.visser@sms.vuw.ac.nz}}
%=================================================================
%=================================================================
%=================================================================
% Simple summary
%\simplesumm{}
%=================================================================
%=================================================================
\abstract{
The nature of coarse graining is intuitively ``obvious'', but it is rather difficult to find explicit and calculable models of the coarse graining process (and the resulting entropy flow) discussed in the literature.  What we would like to have at hand is some explicit and calculable process that takes an arbitrary system, with specified initial entropy $S$, and monotonically and controllably drives the entropy to its maximum value. This does not have to be a physical process, in fact for some purposes it is better to deal with a \emph{gedanken-process}, since then it is more obvious how the ``hidden information'' is hiding in the fine-grain correlations that one is simply agreeing not to look at. We shall present several simple mathematically well-defined and easy to work with conceptual models for coarse graining. We shall consider both the classical Shannon and quantum von Neumann entropies, including models based on quantum decoherence, and analyze the entropy flow in some detail. 
When coarse-graining the quantum von Neumann entropy, we find it extremely useful to introduce an adaptation of Hawking's super-scattering matrix.
These explicit models that we shall construct allow us to quantify and keep clear track of the entropy that appears when coarse-graining the system,  and the information that can be hidden in unobserved correlations.  (While not the focus of the current article, in the longer run these considerations are of interest when addressing black hole information puzzle.)


\bigskip
\noindent
{\sc PACs}:   89.70.Cf; 05.20.-y; 05.30.-d; 05.70.-a;

\bigskip
\noindent
{\sc Published as}:   Entropy {\bf 19:5} (2017) 207; 
(Special Issue: Black Hole Thermodynamics II.)
}

%=================================================================
%=================================================================
%=================================================================
\maketitle
%=================================================================	
%=================================================================	
%=================================================================	
\section{Introduction}
%=================================================================

In this article we will aim to clarify some issues related with the ``coarse graining'' process that is typically asserted to be associated with the monotone increase of entropy implied by the second law of thermodynamics. While many sources discuss the  ``coarse graining'' process in a somewhat intuitive manner, it is extremely difficult to find explicit, controllable, and fully calculable models, see for instance references~\cite{Ridderbos:2002,Gell-Mann:2006,Kofler:2007,Kofler:2008,Simon:2011,Kelly:2013,Visser:2015,Faist:2016}.  (More general background, on coarse-graining in various renormalization group settings, can be found in references~\cite{Wilson:1971a,Wilson:1971b,Wilson:1973,Kadanoff:1976,White:1992,White:1993, Vidal:2002,Hallberg:2006,Vidal:2007}.) What we would be seeking in this current article is some process, (either purely mathematical or possibly physical), that takes an arbitrary system with specified initial entropy $S$, and monotonically drives the entropy to its maximum value --- that is, $\ln N$, for any finite classical or quantum system  described by the Shannon or von Neumann entropies~\cite{Shannon1, Shannon2, von-neumann}. Whenever possible it would be desirable to introduce an adjustable smearing parameter to continuously interpolate between the original system and the smeared system.

For some purposes it is better to deal with a \emph{gedanken-process}, since a \emph{gedanken-process} is in principle reversible, and then it is more obvious how the ``hidden information'' is hiding in the correlations.
The techniques that work best for classical Shannon entropy often do not carry over to quantum von~Neumann entropy, 
or \emph{vice versa}.  We shall develop a number of different and complementary models for coarse graining, adapted to either classical or quantum behaviour, and (classically) to the continuum \emph{versus} discretium character of the system. (The discretium is often alternatively referred to as the discretum; the choice is a matter of taste. Discretum is more Latinate, discretium is more Anglophone.)

The relevance of having explicit, controllable, and calculable methods available to describe coarse-graining emerges when we deal with physical issues of information; specifically the ``hidden information'' lurking in the correlations rendered unobservable by the coarse graining. The current study provides some tools to treat these processes,  which we plan to adapt to understand what is happening with the quantum information in black hole formation and evaporation, where the classical gravitational collapse that initially forms the black hole is an extremely dramatic coarse-graining process; one that encodes ideas about a possible resolution of the black hole information puzzle~\cite{info-puzzle,Hawking:1974,Hawking:1975,Mathur:2008,Mathur:2009,Alonso-Serrano:2015a,Alonso-Serrano:2015b,Alonso-Serrano:2016,Alonso-Serrano:2017}.

%=================================================================
\section{Structure of the article}
%=================================================================

We shall first present a brief historical survey of the coarse graining problem in section \ref{S:history}, and then discuss coarse graining in the context of classical Shannon entropy in section \ref{S:shannon}. 
For Shannon entropy in the continuum a coarse-graining in terms of an idealized and generalized diffusion process, (possibly a \emph{gedanken-process} rather than real physical diffusion), is the easiest to implement.
More generally, for Shannon entropy the distinction between  continuum and discretium is important, and any coarse graining that maps the continuum into the discretium is particularly delicate. 
We then turn to the quantum von Neumann entropy in section \ref{S:von-neumann}, elucidating several coarse-graining processes based on maximal mixing, on partial traces, and on decoherence. We introduce Hawking's super-scattering operator in section \ref{S:super-scattering}, with an aim to using it in section \ref{S:hilbert-diffusion} where we define a wide class of diffusion-like processes on Hilbert space (rather than in physical space). 
Finally we wrap up with a discussion in section \ref{S:discussion}.

\enlargethispage{20pt}	
%=================================================================
\section{History}\label{S:history}
%=================================================================

Coarse-graining plays a crucial role in our description of the world. It allows us to work with very large or complex systems by choosing a more reasonable and tractable number of variables, (and consequently reducing the observable degrees of freedom of the system under consideration).  
See for instance references~\cite{Ridderbos:2002,Gell-Mann:2006,Kofler:2007,Kofler:2008,Simon:2011,Kelly:2013,Visser:2015,Faist:2016}. Coarse-graining should be chosen so as to still describe the key properties we are most interested in, ``integrating over'' properties of percieved lesser importance. These coarse-graining processes are used in very different fields, (from biology to physics to astronomy), but for the current article 
we are mostly interested applications to quantum information theory.  (Though in future we plan to consider quantum fields in curved spacetimes.)

The price of coarse-graining is that some properties or physical effects, corresponding to the degrees of freedom that we are not looking at, will be ``hidden'' in the system. When calculating the entropy of these coarse-grained systems there is an associated increase of system entropy due to uncertainty regarding the coarse-grained degrees of freedom. We shall see that the  difference of entropy after imposing coarse-graining can be understood as ``hidden information'':

\begin{equation}
I_\mathrm{hidden} =S_{\mathrm{coarse}\hbox{-}\mathrm{grained}}
-S_{\mathrm{before}\hbox{-}\mathrm{coarse}\hbox{-}\mathrm{ graining}}.
\end{equation}
Because coarse-graining can sometimes be viewed as a \emph{gedanken-process}, an agreement to simply ``not look at'' certain features of the system, the coarse-grained entropy is unavoidably \emph{contextual} and may be \emph{observer dependent}. 

\enlargethispage{20pt}	

Classically, the way to describe entropy associated with information is provided by the Shannon entropy, which first introduced the interpretation of entropy as a measure of lack of information~\cite{Shannon1,Shannon2}. In the language of communications engineers:

\begin{equation}
S=-\sum_{i=1}^N p_i \log_2 p_i,
\end{equation}
where $p_i$ is the probability associated with each microstate and the entropy is measured in ``shannons''/``bits''. Physicists prefer to use the natural logarithm, $\ln$, and measure the entropy in ``natural units'' (nats). 
There are several ways to define information, though in general, the information can be considered as the reduction of uncertainty. The maximum information, can be defined as the difference between the maximum possible entropy and the value we measure

\begin{equation}
I_\mathrm{max} =S_\mathrm{max}-S = \ln N - S.
\end{equation}
This is often called negentropy~\cite{Shannon1,Shannon2} and it is a measure of distance to complete uncertainty, $p_i = {1\over N}$. 
In the context of quantum resource theory, this can be interpreted as  the unique measure of information that is stable under a subset of ``noisy operations''~\cite{Horodecki:2003}.
In terms of our hidden information due to coarse graining 

\begin{equation}
I_\mathrm{hidden} \in [0, I_\mathrm{max} ].
\end{equation}

In the realm of quantum mechanics, the previous picture changes because the information is stored in a different way and the states can be entangled. The operator that represents the statistics of the physical states can be represented as a density matrix. The analogous entropy now refers to the entanglement entropy between different subsystems that were coarse-grained, and it is formulated by the von Neumann entropy~\cite{von-neumann}:

\begin{equation}
S=-\tr (\rho \ln \rho).
\end{equation}
Here $\rho$ is the density matrix, and $\tr$ denotes the trace.
This von Neumann entropy has to be conserved in unitary quantum evolution. However, there exists the possibility, (or \emph{necessity} in some cases), that we decide to ``look'' only at one part of the system and forget about the rest. This process is different when quantum properties have to be taken into account, and will often correspond to ``trace'' over a specified subsystem. In this quantum case it can also be seen that coarse-graining corresponds to an increment of the entropy completely related with the ``hidden'' quantum information. 

Other interesting questions emerge when the transition from quantum to classical systems is taken into account. Despite valiant efforts this process, the emergence of classicality and its interpretation, is still not completely well-understood. One of the strong possibilities comes from the idea of {\it decoherence}~\cite{Zeh:1986,Zurek:1992,Zurek:1996,Zurek:2003,Joos:2003,Schlosshauer:2007,Halliwell:1999}, that is based on the separation of {\it relevant} and {\it irrelevant} degrees of freedom of the system, and removing the interference terms (off-diagonal terms of the density matrix), that precisely comes from the interaction of the emergent classical system (relevant degrees of freedom) with the environment (irrelevant degree of freedom). It is easy to see that this process arises from a specific process for coarse-graining the system. 

So in this approximation, the classical realm corresponds to a set of coarse-grained non-interfering histories, (called decoherent histories approach to quantum mechanics) \cite{Halliwell:1999,Gell-Mann:2006}. It is important to note that there also exist other interesting ideas about classical emergence, that also involves some notion of coarse-graining. (See, for instance, references~\cite{Kofler:2007,Kofler:2008}.)
	
In both cases, classical and quantum, it is important to understand that the ``hidden information'' corresponds to correlations that we decide not to look at with our particular coarse-graining method.  This implies that the measured entropy depends on the choice of the particular coarse-graining developed, entropy is \emph{contextual} and can be \emph{observer dependent}. For this reason, it is interesting for the physical interpretation to have some knowledge of ``where'' the hidden information is hidden, and  ``what'' we hide, and how we hide it, in order to have complete control of our system. The importance of understanding these  processes can be understood for instance in the study of some open questions related with information, (as for instance the black hole information puzzle~\cite{info-puzzle,Hawking:1974,Hawking:1975,Mathur:2008,Mathur:2009,Alonso-Serrano:2015a,Alonso-Serrano:2015b,Alonso-Serrano:2016,Alonso-Serrano:2017}, where it is not completely well understood where the information flows during the Hawking evaporation process).
In previous literature, coarse graining has been widely considered, (for instance, in references~\cite{Ridderbos:2002,Gell-Mann:2006,Kofler:2007,Kofler:2008,Simon:2011,Kelly:2013,Visser:2015,Faist:2016} and~\cite{Takayanagi:2010,Nieuwenhuizen:2005,Nambu:2009,Krishnan:2017,DiMatteo:2017,Frenzel:2016,Rovelli:2014,Noorbala:2014,Weilenmann:2016}), but in most of those cases coarse-graining is assumed \emph{ab initio}, preventing one from having a complete control of the system. In this paper we shall study different quantifiable models of coarse-graining the entropy, and the selection of a suitable tuneable smearing parameter that will allow us to manipulate coarse-graining in a controllable manner, and understand what is hidden and what is correlated.

%=================================================================
\section{Coarse graining the classical Shannon entropy}\label{S:shannon}
%=================================================================

For a classical system with a finite, or at worst denumerable, number of states (a discretium of states) the Shannon entropy can be defined as

\begin{equation}
S = - \sum p_i \ln p_i; \qquad\qquad  \sum p_i = 1.
\end{equation}
There are many abstract theorems that can be proved for this quantity, but which will not be the main focus here~\cite{Jaynes:1957a,Jaynes:1957b,zipf,shannon}. 
In contrast for a non denumerable number of states (a continuum of states; say 3-space for definiteness) it is better to write

\begin{equation}
S(\rho;\rho_*) = -\int \rho(x) \;\ln\left\{\rho(x)\over\rho_*\right\}\; d^3 x; \qquad\qquad \int \rho(x) \; d^3 x = 1.
\end{equation}
Here $\rho_*$ is some arbitrary-but-fixed parameter that is needed for dimensional consistency; it is  just a convenient normalizing constant that effectively sets the zero for $S$. (In particular, continuum entropy \emph{differences} are always independent of $\rho_*$.) Note that under re-scaling we have the trivial result

\begin{equation}
S(\rho;\lambda\rho_*)=S(\rho;\rho_*) + \ln\lambda.
\end{equation}

\clearpage
Below we exhibit 3 forms of classical coarse graining suitable for three situations:
\begin{itemize}
\item Continuum $\longrightarrow$ continuum; \quad(independent of $\rho_*$).
\item Continuum $\longrightarrow$ discretium;%{\color{red}\footnote{The word discretium is a commonly used minor variant of ``discretum''.}}
\quad (not manifestly independent of $\rho_*$; 
care and discretion required).
\item Discretium $\longrightarrow$ discretium; \quad\,(trivially independent of $\rho_*$).
\end{itemize}

%-----------------------------------------------------------------------------------------------------
\subsection{Continuum $\longrightarrow$ continuum: Diffusion-based forms of coarse graining}
%-----------------------------------------------------------------------------------------------------

Starting from the probability density $\rho(x)$, coarse-grain using some sort of convolution process

\begin{equation}
\rho(x) \to \rho_t(x) = \int K(x,y;t) \;\rho(y)\; d^3y;  \qquad \int K(x,y;t)\; d^3y = 1; \qquad  \int \dot K(x,y;t)\; d^3y = 0.
\end{equation}
Here $K(x,y;t)$ is some positive normalized kernel; $t$ is not necessarily physical time, it is simply a parameter that controls how far you smear out the probability distribution. This kernel can be viewed as representing a ``defocussing'' process. A delta-function initially concentrated at $x_0$, that is $\delta(x-x_0)$, becomes ``smeared out'' to a time-dependent and space-dependent ``blob'' $K(x,x_0,t)$. 
Now from the basic definition of Shannon entropy (in the continuum) we have

\begin{equation}
S_t = -\int \rho_t(x) \ln\left\{\rho_t(x)\over\rho_*\right\} \; d^3x;
\qquad
\dot S_t = -\int \dot \rho_t(x) \ln\left\{\rho_t(x)\over\rho_*\right\} \; d^3x.
\end{equation}

%\clearpage\noindent
Even before specializing the kernel $K$ we see

\begin{equation}
\dot S_t = -\int\int  \dot K(x,y;t) \;\rho(y) \; \ln\left\{\rho_t(x)\over\rho_*\right\} \; d^3y \; d^3 x.
\end{equation}
Here, thanks to the normalizing condition,  the RHS is actually independent of $\rho_*$.

\bigskip
\noindent
Now let us be more specific: Let $K$ be an ordinary diffusion operator, (that is, formally we can write $K = \exp\{t \sigma \nabla^2\}$), where $\sigma$ is the diffusion constant. Then

\begin{equation}
\dot K = \sigma\, \nabla^2 K;
\qquad
\hbox{whence}
\qquad
\dot \rho_t = \sigma\, \nabla^2 \rho_t.
\end{equation}
Therefore, in this ordinary diffusion situation,

\begin{equation}
\dot S_t = -\sigma \int \left(\nabla^2 \rho_t(x)\right) \; \ln\left\{ \rho_t(x)\over\rho_*\right\} \; d^3x.
\end{equation}
Integration by parts (noting that $\rho_*$ drops out of the calculation) now yields

\begin{equation}
\dot S_t = \sigma \int {\nabla \rho_t(x)\cdot \nabla\rho_t(x) \over \rho_t(x)} \; d^3x \geq 0.
\end{equation}
So under diffusion-like-smoothing, entropy always rises as the smearing parameter $t$ increases.
We can easily generalize this to position-dependent, direction-dependent, (and even density-dependent) diffusivity. 
Suppose the kernel (and so the probability density) satisfy

\begin{equation}
\dot K = \nabla_i \left(\sigma^{ij}(x,\rho_t(x)) \nabla_j K\right);
\qquad\qquad
\dot \rho_t = \nabla_i \left(\sigma^{ij}(x,\rho_t(x)) \nabla_j \rho_t\right).
\end{equation}
This now describes a generalized diffusion process, with $\sigma^{ij}(x,\rho(x))$ being a position-dependent, direction-dependent, (and even density-dependent), diffusion ``constant''.
Then under the same logic as previously

\begin{equation}
\dot S_t = \int \sigma^{ij}(x,\rho_t(x)) \; {\nabla_i \rho_t(x)\cdot \nabla_j\rho_t(x) \over \rho_t(x)} \; d^3x \geq 0,
\end{equation}
with non-negativity holding as long as the matrix $\sigma^{ij}(x,\rho_t(x))$ is positive semidefinite.
For either simple or generalized diffusion we can then define

\begin{equation}
I_\mathrm{correlations} = S_t - S_0 \geq 0.
\end{equation}
Here $t$ is not necessarily physical time; it is simply the smearing parameter.
If we view the diffusion process as a \emph{gedanken-process}, a simple de-focussing of our vision described by the smearing parameter $t$, then the effect is clearly reversible, and $I_\mathrm{correlations}$ is simply the information hiding in the correlations that we can no longer see due to our deliberately blurred vision --- but this is information we can easily regain simply by refocussing our vision. Overall, this construction yields a nice explicitly controllable entropy flow as a function of the coarse graining parameter $t$.

%-----------------------------------------------------------------------------------------------------
\subsection{Continuum $\longrightarrow$ discretium: Box-averaging forms of coarse graining}
%-----------------------------------------------------------------------------------------------------

Starting from the continuum, let us now divide the universe up into a denumerable set of non-overlapping boxes $B_i$, that nevertheless cover the entire universe.
Then define individual probabilities $p_i$ for each box,

\begin{equation}
p_i = \int_{B_i} \rho(x) d^3 x,
\end{equation}
so that you are throwing away (agreeing not to look at) detailed information of the probability distribution inside each individual box.
Now compare the continuum and discretium entropies

\begin{equation}
S(\rho,\rho_*) = -\int \rho(x) \ln\left\{\rho(x)\over\rho_*\right\} \; d^3x;
\qquad
\hbox{and}
\qquad
S_B = -\sum p_i \ln p_i.
\end{equation}
We shall consider 2 cases; a ``simple'' all-or-nothing box averaging, and a more subtle box averaging where one first diffuses the probability density in the continuum before implementing box-averaging.

%-----------------------------------------------------------------------------------------------------
\subsubsection{Simple box-averaging}
%-----------------------------------------------------------------------------------------------------

For ``simple'' box averaging, where we fix individual boxes $B_i$ but do not have any tuneable parameter to work with,   we shall show that, (unfortunately the $\rho_*$ dependence is unavoidable):

\begin{equation}
S_B \geq S(\rho,\rho_*) + \hbox{(some $\rho_*$-dependent normalization to set the zero of entropy)}.
\end{equation}
To prove this, use the idea of (classical) relative entropy (a measure of distinguishability between two entropies).  It is a standard result that if $\rho(x)$ and $\rho_0(x)$ are any two normalized probability distributions then

\begin{equation}
\int \rho(x) \ln\left\{\rho(x)\over\rho_0(x)\right\} \, d^3 x \geq 0.
\end{equation}

\clearpage
Let us now define a ``box-wise-constant'' probability density $\rho_B(x)$ as follows

\begin{equation}
\rho_B(x):  \hbox{if $x\in \mathrm{int}(B_i)$ then $\rho_B(x)$ } =   {p_i \over \mathrm{volume}(B_i)}  
= {\int_{B_i} \rho(x) \; d^3x\over \int_{B_i} \; d^3 x}.
\end{equation}
(We do not care what happens on the box boundaries, simply because they are of measure zero.)\\
Then $\rho_B(x)$ is certainly a normalized probability distribution, $\int \rho_B(x) \; d^3 x = 1$.
Consequently

\begin{equation}
\int \rho(x) \ln\left\{\rho(x)\over\rho_B(x)\right\}\, d^3 x \geq 0.
\end{equation}
Whence

\begin{equation}
\int \rho(x) \ln\left\{\rho(x)\over\rho_*\right\}\, d^3 x - \int \rho(x) \ln\left\{\rho_B(x)\over\rho_*\right\}\, d^3 x \geq 0.
\end{equation}
This further implies

\begin{equation}
 - \int \rho(x) \ln\left\{\rho_B(x)\over\rho_*\right\}\,d^3 x \geq - \int \rho(x) \ln\left\{\rho(x)\over\rho_*\right\}\, d^3 x.
\end{equation}
But since $\rho_B(x)$ is piecewise constant (in fact, box-wise constant) on the LHS we then have  

\begin{equation}
- \int \rho(x) \ln\left\{\rho_B(x)\over\rho_*\right\} \, d^3 x= - \int \rho_B(x) \ln\left\{\rho_B(x)\over\rho_*\right\}\, d^3 x
= S(\rho_B),
\end{equation}
and we see that $S(\rho_B) \geq S(\rho)$. But we can actually do quite a bit more, since working in a slightly different direction we have

\begin{eqnarray}
 S(\rho_B) &=& - \int \rho_B(x) \ln\left\{\rho_B(x)\over\rho_*\right\}\, d^3 x
 \nonumber\\
 &=&
 - \sum p_i \ln\left( p_i \over\rho_*  \mathrm{volume}(B_i)\right) 
\nonumber\\
&=& 
- \sum_p p_i \ln p_i + \sum p_i \ln[ \rho_*  \mathrm{volume}(B_i)]
\nonumber\\
&=&
 S_B + \sum p_i \ln[ \rho_*  \mathrm{volume}(B_i)].
\end{eqnarray}
Now using the normalization $\sum_i p_i=1$ we have

\begin{equation}
\sum_i p_i \ln[ \rho_*  \mathrm{volume}(B_i)] =  \ln\left[ \rho_* \; \prod_i \mathrm{volume}(B_i)^{p_i}\right].
\end{equation}
But note that we can interpret

\begin{equation}
 \prod_i \mathrm{volume}(B_i)^{p_i} = \overline{V},
\end{equation}
where $\overline{V}$ is just the (probability weighted) \emph{geometric mean} volume of the boxes $B_i$. 
That is, we have

\begin{equation}
S(\rho_B) = S_B + \ln[ \rho_* \overline{V}] \geq S(\rho,\rho_*).
\end{equation}
Rearranging, we have the rigorous inequality:

\begin{equation}
S_B =S(\rho_B,\rho_*) -  \ln[ \rho_* \overline{V}] \geq S(\rho,\rho_*) -  \ln[ \rho_* \overline{V}].
\end{equation}
The RHS is actually independent of $\rho_*$. To verify this note

\begin{equation}
{d\over d\rho_*} \left( \vphantom{\Big|} S(\rho,\rho_*) -   \ln[ \rho_* \overline{V}]  \right)
=  \int {\rho(x)\over \rho_*} d^3 x -  {1\over\rho_*} = {1\over\rho_*}-{1\over\rho_*} = 0.
\end{equation}
So we are free to pick $\rho_*$ to taste. Three particularly useful options are:

\begin{itemize}
\item 
Simply pick $\rho_* = \overline{V}^{-1}$, in which case we have

\begin{equation}
S_B \geq S\left(\rho,{\overline{V}}^{-1}\right). \end{equation}
\item
If all boxes have the same volume $V$ then automatically we have $\overline{V} = V$. In this situation we simply pick $\rho_* = {V}^{-1}$, in which case

\begin{equation}
S_B \geq S\left(\rho,{{V}}^{-1}\right). 
\end{equation}

\item
Alternatively, keep $\rho_*$ arbitrary, and simply live with the fact that the best we can say is

\begin{equation}
S_B \geq S(\rho,\rho_*) -  \ln[ \rho_* \overline{V}].
\end{equation}
\end{itemize}
The central result is always 

\begin{equation}
S_B > S(\rho,\rho_*) + \hbox{(some $\rho_*$-dependent normalization to set the zero of entropy)}.
\end{equation}
Subject to these technical quibbles we can then define

\begin{equation}
I_\mathrm{correlations} = S_B - S(\rho,\rho_*) + \ln[ \rho_*  \overline{V}] \geq 0.
\end{equation}
This is both independent of $\rho_*$ and guaranteed non-negative. 
Physically this represents the ``hidden information'' encoded in the density-density correlations inside each of the boxes $B_i$, correlations that we are simply agreeing not to look into. Note that box-averaging in this context is an all-or-nothing proposition, there is no tuneable parameter to adjust.  

%-----------------------------------------------------------------------------------------------------
\subsubsection{Continuum diffusion followed by box-averaging}
%-----------------------------------------------------------------------------------------------------

Now let us introduce a tuneable parameter $t$, allowing the continuum probability density $\rho(x)$ to evolve under some diffusion process, (either physical or a \emph{gedanken-process}), so that  $\rho(x) \to \rho_t(x)$ before box-averaging. Then in terms of the (now time-dependent) boxed probability density

\begin{equation}
\rho_{B_t}(x):  \hbox{if $x\in \mathrm{int}(B_i)$ then $\rho_{B_t}(x)$ } =   {(p_i)_t \over \mathrm{volume}(B_i)}  
= {\int_{B_i} \rho_t(x) \; d^3x\over \int_{B_i} \; d^3 x},
\end{equation}
we immediately have without any further need of calculation

\begin{equation}
S(\rho_B) \geq S(\rho); \qquad\qquad S(\rho_{B_t}) \geq S(\rho_t).
\end{equation}
But a more subtle result is this:

\begin{equation}
\dot S(\rho_{B_t})  = - \int \dot\rho_{B_t} \ln\left\{\rho_{B_t}\over\rho_*\right\} dx 
= - \int \dot\rho_t \ln\left\{\rho_{B_t}\over \rho_*\right\} dx
\geq - \int \dot\rho_t \ln\left\{\rho_t\over\rho_*\right\} dx = \dot S(\rho_t) \geq 0.
\end{equation}
Here we have used the fact that $\rho_{B_t}$ is box-wise constant as a function of $x$, and the positivity of relative Shannon entropy. 
This argument now guarantees that $S(\rho_{B_t})$ is monotone non-decreasing as a function of the smearing parameter $t$.
If $\overline{V}$ is held fixed, (for example if all boxes are the same size so that $\overline{V}=V$), this furthermore implies $\dot S_{B_t} \geq 0$. 
However in general, because 

\begin{equation}
{d x^{n(t)}\over dt} = {d\over dt}  \exp(n(t) \ln x) = \dot n \; \ln x \; x^n,
\end{equation}
we have

\begin{equation}
\rho_*\dot{\overline{V}} = {d\over dt} \prod (\rho_* V_i)^{p_i} = \sum \dot p_i  \ln(\rho_* V_i) \prod (\rho_* V_i)^{p_i} 
= \left\{ \sum (\dot p_i  \ln(\rho_* V_i) )\right\} \;   \rho_* \overline{V} .
\end{equation}
Thus 

\begin{equation}
\dot{\overline{V}} = \overline{V} \; \left\{ \sum\dot p_i  \ln(\rho_* V_i) \right\},
\end{equation}
which in general is nonzero (but independent of $\rho_*$) without imposing extra conditions. Nevertheless what we do have is already quire powerful: We always have the result that  $S(\rho_{B_t})$ is monotone non-decreasing, and whenever $\dot{\overline{V}} =0$ we even have $S_{B_t}$ monotone non-decreasing as a function of the smearing parameter $t$. We can then define the ``hidden information'' as 

\begin{equation}
I_\mathrm{correlations} = S_{B_t} - S(\rho_t,\rho_*) + \ln[ \rho_*  \overline{V}_t] \geq 0.
\end{equation}
This is both independent of $\rho_*$ and guaranteed non-negative. 
Physically this represents the ``hidden information'' encoded in the now smearing-parameter-dependent density-density correlations inside each of the boxes $B_i$, correlations that we are simply agreeing not to look into. Note that box-averaging in this context thus yields a nice explicitly controllable entropy flow as a function of the coarse graining parameter $t$.

%\clearpage

%-----------------------------------------------------------------------------------------------------
\subsection{Discretium $\longrightarrow$ discretium: Aggregation/averaging forms of coarse graining}
%-----------------------------------------------------------------------------------------------------

Suppose we have a denumerable set of states with probabilities $p_i$ for which we define

\begin{equation}
S = - \sum p_i \ln p_i.
\end{equation}
(This could be done for instance, \emph{after} breaking the universe up into a denumerable set of boxes.)
Let us now consider two ways of ``aggregating'' these probabilities. 

%-----------------------------------------------------------------------------------------------------
\subsubsection{Naive aggregation}
%-----------------------------------------------------------------------------------------------------

Take any 2 states, (any two boxes), say $a$ and $b$, and simply \emph{average} their probabilities, so that 

\begin{equation}
p_{a,\mathrm{new}} = p_{b,\mathrm{new}} = \bar p = {p_a + p_b\over2}.
\end{equation}
Note that this process leaves the number of states (boxes) fixed; it is inadvisable to \emph{merge} the boxes, as that tends ultimately to \emph{decrease} the entropy in a manner that is physically uninteresting.
Then set $p_a = \bar p \pm\epsilon$, and $p_b=\bar p\mp \epsilon$, (where we can always choose signs so that $\epsilon\geq 0$). We have:

\begin{equation}
- p_a\ln p_a  - p_b\ln p_b  = -(\bar p \pm\epsilon)\ln (\bar p \pm\epsilon)-(\bar p \mp\epsilon)\ln(\bar p \mp\epsilon).
\end{equation}
But

\begin{equation}
{d\over d\epsilon} \left[\vphantom{\Big|}
  -(\bar p \pm\epsilon)\ln (\bar p \pm\epsilon)-(\bar p \mp\epsilon)\ln(\bar p \mp\epsilon) \right]
= \mp \ln(\bar p \pm\epsilon) \pm\ln(\bar p \mp\epsilon) = \ln\left( \bar p -\epsilon\over \bar p +\epsilon\right) \leq 0.
\end{equation}
Therefore

\begin{equation}
-(\bar p \pm\epsilon)\ln (\bar p \pm\epsilon)-(\bar p \mp\epsilon)\ln(\bar p \mp\epsilon) \leq
\left[-(\bar p \pm\epsilon)\ln (\bar p \pm\epsilon)-(\bar p \mp\epsilon)\ln(\bar p \mp\epsilon)\right]_{\epsilon=0}.
\end{equation}
That is

\begin{equation}
- p_a\ln p_a  - p_b\ln p_b  \leq - 2 \bar p \ln \bar p. 
\end{equation}
Thus the result of aggregating (averaging) any two states in this manner is always to increase the Shannon entropy:

\begin{equation}
S_\mathrm{aggregated} \geq S_\mathrm{unaggregated}.
\end{equation}
Note that the end point of this aggregation process occurs when all of the $p_i$ are equal, so $p_i = 1/N$, and  $S\to\ln N$, which is its maximum possible value.
We can now define the ``hidden information'' as 

\begin{equation}
I_\mathrm{correlations} = S_\mathrm{aggregated} - S_\mathrm{unaggregated} \geq 0.
\end{equation}
These are now correlations between the states (boxes) that we lumped together in the averaging process.
(This is at this stage an ``all-or-nothing'' process with no tuneable parameter).

%-----------------------------------------------------------------------------------------------------
\subsubsection{Asymmetric aggregation}
%-----------------------------------------------------------------------------------------------------
A generalization of this notion of aggregation is to define an ``asymmetric averaging''

\begin{equation}
p_{a,\mathrm{new}} =  \lambda p_a + (1-\lambda) p_b; \qquad
p_{b,\mathrm{new}} =  (1-\lambda) p_a + \lambda p_b; \qquad \lambda\in(0,1).
\end{equation}
Then $\bar p_{new} = \bar p$, the average is unaffected, whereas $|p_{a,\mathrm{new}}-p_{b,\mathrm{new}}| = |2\lambda-1|\; |p_a-p_b| < |p_a-p_b|$. So the new $p_a$ and $p_b$ move ``closer together''. 

Then we can again write $p_a = \bar p \pm\epsilon$, and $p_b=\bar p\mp \epsilon$, and in addition we have $p_{a,\mathrm{new}} = \bar p \pm\epsilon_2$, and $p_{b,\mathrm{new}}=\bar p\mp \epsilon_2$, with both $\epsilon$ and $\epsilon_2$ positive, and $\epsilon_2<\epsilon$.  But then the same argument as used previously now gives

\begin{equation}
-(\bar p \pm\epsilon)\ln (\bar p \pm\epsilon)-(\bar p \mp\epsilon)\ln(\bar p \mp\epsilon) \leq
\left[-(\bar p \pm\epsilon)\ln (\bar p \pm\epsilon)-(\bar p \mp\epsilon)\ln(\bar p \mp\epsilon)\right]_{\epsilon=\epsilon_2}.
\end{equation}
That is

\begin{equation}
- p_a\ln p_a  - p_b\ln p_b  \leq - p_{a,new}\ln p_{a,new}  - p_{b,new}\ln p_{b,new}  . 
\end{equation}
Thus the result of aggregating (averaging) any two states in this manner is always to increase the Shannon entropy.  
The rest of the argument follows as previously; with the same conclusions.
(This argument can also be rephrased in terms of concavity of the Shannon entropy as a function of the $p_i$.)
We can again define the ``hidden information'' as 

\begin{equation}
I_\mathrm{correlations} = S_\mathrm{aggregated} - S_\mathrm{unaggregated} \geq 0.
\end{equation}
These are now correlations between the states (boxes) that we lumped together in the asymmetric averaging process.

%=================================================================
\subsection{Summary: Coarse graining classical Shannon entropy}
%=================================================================

As we have just seen, it is possible to construct a number of reasonably well-behaved models for classical coarse-graining of the  Shannon entropy. Some of these models are ``all-or-nothing'', without any tuneable smearing parameter, whereas other models depend on a tuneable coarse graining parameter, (that we have chosen to designate $t$ even when it might not represent physical time). The trickiest of these classical constructions is when one coarse-grains a continuum down to a discretium, since the normalizing parameter $\rho_*$, (which is essential to normalizing the continuum Shannon entropy), no longer appears in the discrete Shannon entropy. Which if any of these models is most relevant depends on the particular process under consideration; in all cases these constructions provide a well-defined framework to use, 
and allow explicit definition and determination of the hidden information.

%=================================================================
\section{Coarse graining the quantum von Neumann entropy}\label{S:von-neumann}
%=================================================================

For quantum systems described by a normalized density matrix $\rho$ the von Neumann entropy is 

\begin{equation}
S = - \tr (\rho \ln \rho); \qquad\qquad  \tr(\rho)=1. 
\end{equation}
There is no continuum \emph{versus} discretium distinction to worry about, and since the density matrix is automatically dimensionless, similarly there  is never any need to introduce a ``reference density'' $\rho_*$. There are still at least three interesting and fundamental notions of coarse graining to consider:
\vspace{-5pt}
\begin{itemize}
\itemsep0pt
\item Maximal mixing.
\item Partial traces (4 sub-varieties).
\item Decoherence (2 sub-varieties). 
\end{itemize}
\vspace{-5pt}
In section \ref{S:hilbert-diffusion} we shall subsequently extend the constructions of this current section by the use of a variant of Hawking's super-scattering operator to be introduced in section \ref{S:super-scattering}. 

%-----------------------------------------------------------------------------------------------------
\subsection{Quantum coarse-graining by maximal mixing}
%-----------------------------------------------------------------------------------------------------
\enlargethispage{20pt}
For simplicity let the dimension of the Hilbert space be finite: $N$. Now let $I_N$ denote the $N$-dimensional identity matrix. 
Then ${I_N\over N}$ is a properly normalized density matrix with entropy $\ln N$.  In fact this is the maximally mixed density matrix with maximum entropy given the dimensionality of the Hilbert space.

\enlargethispage{20pt}	
Now consider this \emph{gedanken-process}:
Acting on the density matrix $\rho$, to keep things properly normalized,  take:
\begin{equation}
\rho \to  \rho_s = (1-s)\rho + s \; {I_N\over N}; \qquad s\in[0,1].
\end{equation}
Here one is using the smearing parameter $s$ (which is certainly not physical time) to selectively ignore what would otherwise be physically accessible information; effectively one is ``blurring'', (that is, coarse-graining)  the system by throwing away (ignoring) position information in Hilbert space. (This is strictly a \emph{gedanken-process}, we know of no physical way of implementing this particular mapping. Nevertheless it is still appropriate to call this a [mathematical] coarse graining.)

Now use the fact that the von Neumann entropy is strictly concave in the density matrix

\begin{equation}
S((1 - s)\rho_0 + s\rho_1)  \geq  (1  - s)S(\rho_0) + sS(\rho_1),
\end{equation}
whence

\begin{equation}
S(\rho_s)  \geq  (1  - s)S(\rho_0) + s\ln N.
\end{equation}
That is

\begin{equation}
S(\rho_s)-S(\rho_0)  \geq s \{ \ln N - S(\rho_0)\}.
\end{equation}
Then we can formally define the hidden information by setting

\begin{equation}
I_\mathrm{correlations} = S(\rho_s)-S(\rho_0)  \geq s \{ \ln N - S(\rho_0)\} \geq 0.
\end{equation}
This represents the amount of information that is temporarily lost in the $s$-dependent blurring process, 
which in view of the \emph{gadanken-nature} of the blurring process, is recoverable simply by de-blurring one's vision.
In this construction one also has a strict upper bound on the hidden information

\begin{equation}
I_\mathrm{correlations} = S(\rho_s)-S(\rho_0)  \leq  \{ \ln N - S(\rho_0)\}.
\end{equation}
It is important to note that this partial maximal mixing procedure involves some subtleties in its interpretation. Compared to other more physical coarse-graining procedures, maximal mixing has a more mathematical flavour. Nevertheless it is still appropriate to call it coarse graining due it continues to satisfy the crucial feature of selectively ignoring some physical aspects of the system, which in turn leads to an effective amount of hidden information.

%-----------------------------------------------------------------------------------------------------
\subsection{Quantum coarse-graining by partial traces}
%-----------------------------------------------------------------------------------------------------

Many quantum information sources and texts hint at how taking partial traces is an example of ``coarse graining", but never quite seem to really get explicit about it. Below we outline a few explicit constructions, some tuneable, some ``all-or-nothing''. 

%-----------------------------------------------------------------------------------------------------
\subsubsection{Non-tuneable partial traces}
%-----------------------------------------------------------------------------------------------------
Suppose the Hilbert space factorizes $H_{AB} = H_A \otimes H_B$, (though the density matrix need not factorize), and define ``simultaneous coarse grainings'' onto the sub-spaces $H_A$ and $H_B$ by:

\begin{equation}
\rho_{AB} \to \rho_{A:B} = \rho_A \otimes \rho_B = \tr_B(\rho_{AB}) \otimes  \tr_A(\rho_{AB}).
\end{equation}
Here the notation A:B effectively means this: Slide a (unphysical, purely mathematical) barrier between the sub-spaces $A$ and $B$ so there is no longer any cross-talk between the two sub-spaces. (Notationally it is more common to see notation such as $S_{A;B}$, but this seems to us to unnecessarily suggest an asymmetry between the two factor sub-spaces.)
The idea is that people ``living in'' $A$ agree not to look into $B$ and \emph{vice versa}.
Then it is relatively easy to prove the quite standard result that:

\begin{equation}
S_{A:B} = \tr \left(\vphantom{\Big{|}} [\rho_A \otimes \rho_B] \ln[\rho_A \otimes \rho_B] \right)
= \tr \left(\vphantom{\Big{|}} [\rho_A \otimes \rho_B] \{ \ln[\rho_A] \otimes I_B + I_A\otimes \ln[\rho_B]\} \right)
= S_A + S_B,
\end{equation}
and furthermore to derive the ``triangle inequalities'':
\enlargethispage{20pt}

\begin{equation}
|S_A - S_B| \leq S_{AB} \leq S_A + S_B.
\end{equation}

\noindent
Combining these results we see

\begin{equation}
S_{A:B} = S_A + S_B \geq S_{AB} \geq |S_A - S_B|.
\end{equation}
The hidden information can be defined by

\begin{equation}
I_\mathrm{correlations} = S_{A:B} - S_{AB} = S_A + S_B - S_{AB}.
\end{equation}
This is just the completely usual notion of mutual information.
The triangle inequalities give

\begin{equation}
I_\mathrm{correlations} = S_{A:B} - S_{AB} = S_A + S_B - S_{AB} \geq 0,
\end{equation}
(so the hidden information is always non-negative), and futhermore

\begin{equation}
I_\mathrm{correlations} = S_{A:B} - S_{AB} = S_A + S_B - S_{AB} \leq S_A + S_B  - |S_A - S_B| = 2 \min\{S_A, S_B\},
\end{equation}
so the hidden information is always bounded from above.
Note the current construction has no tuneable parameter; the coarse-graining is ``all-or-nothing'', a feature that we shall now show how to evade.

%-----------------------------------------------------------------------------------------------------
\subsubsection{Tuneable partial traces}\label{sss:tuneable-partial-trace}
%-----------------------------------------------------------------------------------------------------

Let us now consider this model:

\begin{equation}
\label{E:partial-tuning}
\rho_{AB} \to \rho_s = (1-s)\,  \rho_{AB} + s \, \rho_A \otimes \rho_B;  \qquad s\in[0,1].
\end{equation}
This smoothly interpolates between the identity process at $s=0$ and the factorized subspace process described in the previous subsection at $s=1$.  (The parameter $s$ is not physical time; it merely characterizes the amount of coarse graining; so this is at this stage a \emph{gedanken-process}.)
Concavity of the von Neumann entropy yields

\begin{equation}
S(\rho_s) = S\left((1-s)\,  \rho_{AB} + s \, \rho_A \otimes \rho_B\right) \geq (1-s) S(\rho_{AB}) 
+ sS\left(  \rho_A \otimes \rho_B\right),
\end{equation}
whence

\begin{equation}
S(\rho_s) 
\geq 
(1-s) S_{AB} + s (S_A+S_B).
\end{equation}
By the triangle inequality this implies

\begin{equation}
S(\rho_s) 
\geq 
S_{AB} + s (S_A+S_B-S_{AB}) \geq S_{AB}. 
\end{equation}
On the other hand, because partial trace applied to $\rho_s$ yields $\rho_s \to \rho_A\otimes\rho_B$ (independent of the value of $s$ as long as it is in the range $s\in[0,1]$) we have

\begin{equation}
S(\rho_s)  \leq S(\rho_A\otimes\rho_B) = S_A + S_B.
\end{equation}
Now consider the entropy flow under evolution of the parameter $s$:  

\begin{equation}
\dot S(\rho_s) = - \tr(\dot\rho_s \ln\rho_s) =  - \tr([(\rho_A \otimes \rho_B) -\rho_{AB}] \ln\rho_s).
\end{equation}
But we can write this as

\begin{equation}
\dot S(\rho_s) =  +\tr\left(\left[\rho_s - [\rho_A \otimes \rho_B] \over 1-s \right] \ln\rho_s\right)
= \left\{  -\tr([\rho_A \otimes \rho_B] \ln\rho_s) -S(\rho_s)\over 1- s\right\}.
\end{equation}
Now applying the Klein inequality for quantum relative entropies, the positivity of relative von Neumann entropy, we have

\begin{equation}
\tr([\rho_A \otimes \rho_B] \ln\rho_s) \leq \tr([\rho_A \otimes \rho_B] \ln[\rho_A \otimes \rho_B]) = -S(\rho_A)-S(\rho_B).
\end{equation}
Therefore

\begin{equation}
\dot S(\rho_s) \geq {1\over1-s} \{ S(\rho_A)+S(\rho_B) -S(\rho_s)\} \geq 0.
\end{equation}
Thus this tuned entropy is nicely monotonic (non-decreasing) in the tuning parameter $s$, and is a good candidate for a (explicit and controllable) coarse-graining process.
We can then safely define the ``hidden information'' as

\begin{equation}
I_\mathrm{correlations} = S(\rho_s) - S_{AB}.
\end{equation}
There is still a nice lower bound

\begin{equation}
I_\mathrm{correlations} \geq s (S_A+S_B-S_{AB}) \geq 0,
\end{equation}
guaranteeing positivity of the hidden information.
On the other hand, the best general upper bound seems to be relatively weak

\begin{equation}
I_\mathrm{correlations} \leq S(\rho_s) - |S_A-S_B|  \leq S_A + S_B  - |S_A - S_B| = 2 \min\{S_A, S_B\}.
\end{equation}

%-----------------------------------------------------------------------------------------------------
\subsubsection{Asymmetric partial trace and maximal mixing}\label{sss:mixed}
%-----------------------------------------------------------------------------------------------------

Here is a more brutal \emph{asymmetric} combination of partial trace and maximal mixing:

\begin{equation}
\rho_{AB} \to  \rho_A \otimes {I_B\over N_B} = \tr_B(\rho_{AB}) \otimes  {I_B\over N_B}.
\end{equation}
Here one is not just taking a partial trace over the subsystem B, one is also maximally mixing the subspace B.
So this process is \emph{asymmetric} in the way it treats subsystems A and B.
(This is best thought of as a \emph{gedanken-process}.)
To verify that entropy has increased in this ``all-or-nothing'' coarse graining process we simply note that

\begin{equation}
S(\rho_{AB}) \leq  S(\rho_A \otimes \rho_B) \leq S\left(\rho_A \otimes {I_B\over N_B}\right).
\end{equation}
Where, specifically, it is easy to see that

\begin{equation}
S\left(\rho_A \otimes {I_B\over N_B}\right) = S_A + \ln N_B.
\end{equation}
The hidden information can easily be defined by 

\enlargethispage{20pt}	
\begin{equation}
I_\mathrm{correlations} = S\left(\rho_A \otimes {I_B\over N_B}\right) - S_{AB} = S_A + \ln N_B - S_{AB} \geq \ln N_B - S_B\geq 0,
\end{equation}
and is guaranteed non-negative.

There is also an upper bound
\begin{equation}
I_\mathrm{correlations} = S_A + \ln N_B - S_{AB} \leq S_A + \ln N_B - |S_A-S_B| \leq (\ln N_B-S_B) + 2 \min\{S_A,S_B\} .
\end{equation}
This seems to be the best upper bound one can extract in this context.

%-----------------------------------------------------------------------------------------------------
\subsubsection{Asymmetric partial trace plus maximal mixing, with tuneable parameter}
%-----------------------------------------------------------------------------------------------------

Starting from the previous case, now consider the effect of adding a tuneable parameter:
\begin{equation}
\rho_{AB} \to \rho_s = (1-s) \rho_{AB} + s \rho_A \otimes {I_B\over N_B}; \qquad s\in[0,1].
\end{equation}
(This is again best thought of as a \emph{gedanken-process}, rather than a physical coarse graining.)
By concavity of the von Neumann entropy

\begin{equation}
S(\rho_s) 
= S\left((1-s)\,  \rho_{AB} + s \, \rho_A \otimes {I_B\over N_B}\right) 
\geq (1-s) S(\rho_{AB})    + sS\left(  \rho_A \otimes {I_B\over N_B}\right),
\end{equation}
so that

\begin{equation}
S(\rho_s) 
\geq 
(1-s) S_{AB} + s (S_A+\ln N_B) = S_{AB} + s \{ S_A+\ln N_B - S_{AB} \} \geq S_{AB}.
\end{equation}
To establish an upper bound note that under the process envisaged in the previous subsection \ref{sss:mixed} (asymmetric partial trace and mixing) we have (now for all $s$ in the appropriate interval)

\begin{equation}
 \rho_s \to \rho_A \otimes {I_B\over N_B} 
\end{equation}
so by the arguments in the previous subsection \ref{sss:mixed} we have

\begin{equation}
S(\rho_s) \leq S_A+\ln N_B.
\end{equation}
Checking monotonicity in the parameter $s$ now requires that we investigate

\begin{equation}
\dot S(\rho_s) = - \tr(\dot\rho_s\ln\rho_s) = - \tr\left( \left[  \rho_A \otimes {I_B\over N_B}  - \rho_{AB} \right] \ln \rho_s \right)
\end{equation}
To proceed we adapt the discussion in subsection \ref{sss:tuneable-partial-trace} (tuneable partial traces). 
We note we can write the above as 

\begin{equation}
\dot S(\rho_s) =  +\tr\left(\left[\rho_s - [\rho_A \otimes I_B/N_B] \over 1-s \right] \ln\rho_s\right)
= \left\{  -\tr([\rho_A \otimes I_B/N_B] \ln\rho_s) -S(\rho_s)\over 1- s\right\}
\end{equation}
Now applying the Klein inequality, the positivity of relative von Neumann entropy, we have

\begin{equation}
\tr([\rho_A \otimes I_B/N_B] \ln\rho_s) \leq \tr([\rho_A \otimes I_B/N_B] \ln[\rho_A \otimes I_B/N_B]) = -S(\rho_A)-\ln N_B.
\end{equation}
Therefore

\enlargethispage{20pt}	
\begin{equation}
\dot S(\rho_s) \geq {1\over1-s} \{ S(\rho_A)+\ln N_B -S(\rho_s)\} \geq 0.
\end{equation}
Thus this partial trace plus maximal mixing tuneable entropy is nicely monotonic (non-decreasing) in the tuning parameter $s$, and is another good candidate for a (explicit and controllable) coarse-graining process.
Then for the ``hidden information'' we can safely define
\begin{equation}
I_\mathrm{correlations} = S(\rho_s) - S_{AB}.
\end{equation}

We note
\begin{equation}
I_\mathrm{correlations} \geq s (S_A+\ln N_B-S_{AB}) \geq s(\ln N_B-S_B) \geq 0,
\end{equation}
thereby guaranteeing non-negativity and a nice lower bound. The best general upper bound in this context seems to be
\enlargethispage{20pt}

\begin{equation}
I_\mathrm{correlations} \leq S(\rho_s) - |S_A-S_B| \leq  
S_A + \ln N_B - |S_A-S_B| .
\end{equation}

%-----------------------------------------------------------------------------------------------------
\subsection{Quantum coarse-graining defined by decoherence}
%-----------------------------------------------------------------------------------------------------
 Choose any fixed but arbitrary basis for the Hilbert space, and let $P_a$ be the corresponding one-dimensional
 projection operators onto these basis elements. So we have
 
    \begin{equation}
    P_a P_b = \delta_{ab} P_b;  \qquad P_a^2 = P_a;  \qquad \sum_a P_a = I_N.
    \end{equation}
Now use these projection operators to define decoherence (\emph{relative to the specified basis}).    
Decoherence is typically viewed as a physical process associated with a semi-classical environment associated with the specified basis, but we could also think of it as a \emph{gedanken-process} where we are just agreeing to ignore part of the density matrix.  (For general background on decoherence see for instance references~\cite{Zeh:1986, Zurek:1992, Zurek:1996, Zurek:2003,Joos:2003,Schlosshauer:2007,Halliwell:1999}.)

%-----------------------------------------------------------------------------------------------------
\subsubsection{Full decoherence}\label{sss:decohere}
%-----------------------------------------------------------------------------------------------------    
    Define full decoherence as follows:     
    
    \begin{equation}
    \rho \to  \rho_D =  \sum_a P_a \; \tr (P_a \rho).
    \end{equation}
 (This preserves the $\tr(\rho)=1$ condition, and kills off all non-diagonal elements in the density matrix.)
 When viewed as a \emph{gedanken-process} one is effectively agreeing to not look at the off-diagonal elements.
 Now in this particular case concavity does not tell one anything useful:

 \begin{equation}
S(\rho_D) = S\left( \sum_a P_a \; \tr (P_a \rho)\right) \geq  \sum_a \tr (P_a \rho) S(P_a) = 0.
\end{equation}
One just obtains the trivial bound $S(\rho_D)\geq 0$. 
Considerably more interesting in this context is the relative entropy inequality, (the Klein inequality), which states that for any two density matrices $\rho$ and $\sigma$: 

\begin{equation}
\tr( \rho [\ln \rho - \ln\sigma]) \geq 0.
\end{equation}
In particular 

\begin{equation}
\tr( \rho [\ln \rho - \ln\rho_D]) \geq 0,
\end{equation}
which can be rearranged to yield

\begin{equation}
-\tr( \rho\ln\rho_D) \geq  - \tr( \rho \ln \rho) = S(\rho).
\end{equation}
Now $\rho_D$ is diagonal, so $\ln\rho_D$ is diagonal, so in the trace in the LHS above only the diagonal elements of $\rho$ contribute:
\begin{equation}
-\tr( \rho\ln\rho_D) = -\tr( \rho_D\ln\rho_D) = S(\rho_D).
\end{equation}

So overall we see that $S(\rho_D)\geq S(\rho)$, the decoherence process is entropy non-decreasing, (as it should be). 
In a completely standard manner we now define the hidden information as

\begin{equation}
I_\mathrm{correlations} =  S(\rho_D) - S(\rho) \geq 0,
\end{equation}
where the hidden information is hiding in the off-diagonal correlations that one has agreed to not look at in implementing the decoherence process. The best upper bound seems to be
\begin{equation}
I_\mathrm{correlations} =  S(\rho_D) - S(\rho) \leq \ln N - S(\rho).
\end{equation}
The full decoherence of the off-diagonal terms allow us to completely forget about quantum effects; the coarse grained system is effectively classical, (albeit a classical stochastic system). 

%\clearpage
%-----------------------------------------------------------------------------------------------------
\subsubsection{Partial (tuneable) decoherence}
%-----------------------------------------------------------------------------------------------------    
 
Now define partial decoherence as follows:     
    
 \begin{equation}
 \rho \to   \rho_{D_s} =  (1-s) \rho + s \rho_D = (1-s) \rho + s \left\{ \sum_a P_a \tr (P_a \rho) \right\}.
 \end{equation}
 (This preserves the $\tr(\rho)=1$ condition, but now only partially suppresses all non-diagonal elements in the density matrix.)
 We note that concavity implies
 
\begin{equation}
S(\rho_{Ds}) = S\left(  (1-s) \rho + s \rho_D\right) \geq (1-s) S(\rho) + s S(\rho_D) 
= S(\rho) + s\{ S(\rho_D) - S(\rho)\} \geq S(\rho).
\end{equation}
By using the Klein inequality

\begin{equation}
\tr(\rho_{D_s}\ln\rho_{D_s}) \leq \tr(\rho_{D_s}\ln\rho_{D})
\end{equation}
But since in the current situation $\rho_D$ is diagonal, (and so $\ln\rho_D$ is also diagonal), and since all the diagonal elements of $\rho_{D_s}$ are by construction constants independent of $s$, this implies

\begin{equation}
\tr(\rho_{D_s}\ln\rho_{D_s}) \leq \tr(\rho_{D}\ln\rho_{D})
\end{equation}
Thus $S(\rho_D)\geq S(\rho_{D_s})$, which has the nice interpretation that partial decoherence involves less entropy than full decoherence. Overall we obtain the chain of inequalities

\begin{equation}
S(\rho) = S(\rho_{D_0}) \leq S(\rho_{D_s})  \leq S(\rho_{D_1}) = S(\rho_D).
\end{equation}
We can do even more, and demonstrate that under partial decoherence the entropy increases monotonically in the control parameter $s$. We note

\begin{eqnarray}
\dot S(\rho_{D_s}) &=& - \tr( \dot \rho_{D_s}\ln\rho_{D_s}) 
\nonumber\\
&=&- \tr( [\rho_D-\rho]\ln\rho_{D_s}) 
\nonumber\\
&=&-  \tr\left( \left[{\rho_D-\rho_{D_s}\over 1-s}\right]\ln(\rho_{D_s})\right) 
\nonumber\\
&=& {1\over 1-s} \left( -S(\rho_{D_s})  - \tr( \rho_D \ln(\rho_{D_s}))     \right).
\end{eqnarray}
Again appealing to the Klein inequality, we have $\tr( \rho_D \ln(\rho_{D_s}))  \leq \tr( \rho_D \ln(\rho_{D}))  = - S(\rho_D)$, 
whence 

\begin{equation}
\dot S(\rho_{D_s})  \geq {S(\rho_{D}) - S(\rho_{D_s}) \over 1-s} \geq 0.
\end{equation}
So again we have a nice calculable monotonically non-decreasing explicit  model for coarse graining.
In a completely standard manner we now define the hidden information as

\begin{equation}
I_\mathrm{correlations} =  S(\rho_{D_s}) - S(\rho) \geq s\{ S(\rho_D) - S(\rho)\}\geq 0,
\end{equation}
where the hidden information is hiding in the off-diagonal correlations that have been partially suppressed because one has agreed to not look at them too carefully in implementing the decoherence process.
The best generic upper bound seems to be 

\begin{equation}
I_\mathrm{correlations} =  S(\rho_{D_s}) - S(\rho) \leq \ln N- S(\rho).
\end{equation}
Thus we have obtained the same upper bound as in the previous subsection, where we completely suppressed the off-diagonal terms.

%=================================================================
\subsection{Summary: Coarse graining quantum von Neumann entropy}
%=================================================================

In the current section we have built three distinct explicit mathematical/physical models of quantum coarse-graining of the von~Neumann entropy. These models were based on maximal mixing; on partial traces (4 sub-varieties); and on decoherence (2 sub-varieties), 7 sub-varieties in total. Some of these models were ``all-or-nothing'' (untuneable) models; others have an explicit coarse graining parameter $s$, and for these tuneable models we carefully proved that the entropy was monotonic in the smearing parameter. This already gives one a much more concrete idea of what coarse graining in a quantum context should be taken to be. In the next two sections we shall extend these ideas first by showing that these coarse graining processes can be viewed as specific examples of Hawking's super-scattering operator~\cite{info-puzzle}; and then apply super-scattering to define a notion of diffusion in Hilbert space.

%=================================================================
\section{Hawking's super-scattering operator}\label{S:super-scattering}
%=================================================================	

%-------------------------------------------------------------------------------------------------------------------------------------------
\def\lint{\hbox{\Large $\displaystyle\int$}}   %needs \usepackage{amssymb} [large integral]
\def\hint{\hbox{\huge $\displaystyle\int$}}  %needs \usepackage{amssymb} [huge integral]
%------------------------------------------------------------------------------------------------------------------------------------------
\def\ss{\hbox{\large $\displaystyle\$ $}}
%------------------------------------------------------------------------------------------------------------------------------------------

In an attempt to develop a non-unitary version of quantum mechanics Hawking introduced the notion of a super-scattering operator  $\ss$, see reference~\cite{info-puzzle}. At its most basic, a super-scattering operator is taken to be a trace-preserving linear mapping from density matrices to density matrices: $\rho \to \ss\rho$, with $\tr(\rho)=1$ implying $\tr(\ss\rho)=1$.  A \emph{proper} super-scattering operator is one that \emph{cannot} simply be factorized into a tensor product of unitary scattering matrices $\mathbb{S}$. That is $\ss \neq \mathbb{S}^\dagger \otimes \mathbb{S}$, implying   $\rho \to \ss\rho \neq \mathbb{S}^\dagger \,\rho \,\mathbb{S}$. 
While these ideas were originally mooted in a black hole context, there is nothing intrinsically black hole related in the underlying concept.  It is sometimes implicitly demanded that super-scattering operators be entropy non-decreasing $S(\ss\rho) \geq S(\rho)$, but one could just as easily demand linearity and trace-preservation as the key defining characteristics, and then divide super-scattering operators into entropy non-decreasing, entropy non-increasing, and other. This point of view has the advantage that the inverse of a super-scattering operator $\ss^{-1}$, \emph{if it exists}, is also a super-scattering operator.

\enlargethispage{20pt}	
To add to the confusion, depending on the sub-field, people often use \emph{different terminology} for essentially the same concept, such as ``trace-preserving (completely) positive operators'', or ``quantum maps'', or ``quantum process'', or ``quantum channels''. Usage (and precise definition) within various sub-fields is (unfortunately) \emph{not entirely standardized}. We will try to keep things simple with a minimal set of assumptions, and with terminology that is likely to resonate widely with the larger sub-communities of interest.

Perhaps the simplest super-scattering process one can think of is this:
Pick a set of positive numbers $p_i$ such that $\sum_i p_i=1$ (which you might like to think of as classical probabilities) and a set of unitary matrices $U_i$. Now consider the mapping

\begin{equation}
\rho \to  \ss_{\{p,U\}} \, \rho = \rho_{\{p,U\}} =  \sum_i p_i \; U_i^\dagger \,\rho\, U_i.
\end{equation}
This is not a unitary transformation, it is instead an ``incoherent sum of unitary transformations''.  But note that the process is linear and trace preserving. 
Then for this particular super-scattering process we have

\begin{equation}
S_{\{p,U\}} = \tr( \rho_{\{p,U\}} \ln \rho_{\{p,U\}} ) = 
\tr\left( \left[ \sum_i p_i \;U_i^\dagger \,\rho\, U_i \right] \ln\left[ \sum_i p_i \; U_i^\dagger\, \rho\, U_i\right] \right) .
\end{equation}
By concavity of the von Neumann entropy

\begin{equation}
S_{\{p,U\}} = \tr( \rho_{\{p,U\}} \ln \rho_{\{p,U\}} ) \geq \sum_ i p_i\; \tr\left( \left[U_i^\dagger\, \rho\, U_i \right] \ln\left[U_i^\dagger\, \rho\, U_i\right] \right)  =  \sum_ i p_i \; \tr(\rho\ln\rho) = S(\rho) = S.
\end{equation}
That is, this particular super-scattering operator  is entropy non-decreasing, $S_{\{p,U\}} \geq S$. Unfortunately we do not have any nice monotonicity result describing how $S_{\{p,U\}}$ grows as a function of the $p_i$ and the $U_i$, this is again an ``all-or-nothing scenario''. (Indeed, systematically finding constructions for which we are always guaranteed a nice monotonicity result is the topic of the next section.) 

As an aside we point out that if one chooses the $p_i$ and $U_i$ from some classical stochastic ensemble then one should work with (classical) expectation values: 

\begin{equation}
\rho \to  \ss_{\langle p,U\rangle} \, \rho =  \left\langle \sum_i p_i \; U_i^\dagger \,\rho\, U_i \right\rangle.
\end{equation}
A suitable choice of (classical) statistical ensemble for the $p_i$ and $U_i$ could then be used to build the decoherence operator $\rho\to \rho_D$ described in subsection \ref{sss:decohere} above.

In a rather different direction, many of the coarse graining processes considered in section \ref{S:von-neumann} above can be reinterpreted as super-scattering processes. 
For instance, in addition to the construction given immediately above, the following are all entropy non-decreasing super-scattering processes:
%============================================================
{\small
\begin{center}
\begin{tabular}{|| l || l || c || }
\hline\hline
Name & Definition & Condition \\ 
\hline\hline
Maximal &   $\vphantom{\Bigg|}\displaystyle{\rho\to \ss_M \,\rho  = (1-s)\rho + s \; {I_N\over N}}$ & $s\in[0,1]$. \\
\hline
Partial/maximal & $\vphantom{\Bigg|}\displaystyle{\rho\to \ss_P \,\rho = \rho_A \otimes {I_B\over N_B}}$ & --- \\
\hline
Partial/maximal/tuneable &  $\vphantom{\Bigg|}\displaystyle{\rho\to \ss_{P_s} \,\rho = (1-s) \rho + s\rho_A \otimes {I_B\over N_B}}$& $s\in[0,1]$. \\
\hline
Full decoherence & $\vphantom{\Bigg|}\displaystyle{\rho\to \ss_D \,\rho = \sum_a P_a \; \tr (P_a \rho)}$& --- \\
\hline
Partial decoherence &$\vphantom{\Bigg|}\displaystyle{\rho\to \ss_{D_s} \,\rho = (1-s) \rho + s \left\{ \sum_a P_a \tr (P_a \rho) \right\}}$ & $s\in[0,1]$. \\
\hline
Incoherent sum &  $\vphantom{\Bigg|}\displaystyle{\rho\to \ss_{\{p,U\}} \,\rho = \sum_i p_i \; U_i^\dagger \,\rho\, U_i}$ & --- \\
\hline\hline
\end{tabular}
\end{center}
}
%========================================================
Oddly enough it is the partial trace process $\rho_{AB} \to \rho_A \otimes \rho_B$ which is not a super-scattering process.  The problem is that while the individual partial traces $\rho_{AB} \to \rho_A$ and $\rho_{AB} \to \rho_B$ are both linear functions; the direct-product reassembly process $\rho_A \otimes \rho_B$ is \emph{not} linear.
So $\rho_{AB} \to \rho_A \otimes \rho_B$ is not a linear mapping; it is not a super-scattering process.
However, when viewed as a \emph{nonlinear} process   $\rho_{AB} \to \mathcal{N} \rho_{AB} = \rho_A \otimes \rho_B$, the nonlinear operator $\mathcal{N}$ is idempotent, $\mathcal{N}^2=\mathcal{N}$. We will come back to this point later.

%=================================================================
\section{Diffusion on Hilbert space}\label{S:hilbert-diffusion}
%=================================================================	

\enlargethispage{20pt}
We shall now develop a notion of diffusion on Hilbert space, similar to what we did in physical space when considering continuum Shannon entropies. There are two routes to consider:
\begin{itemize}
\item Diffusion via super-scattering.
\item Diffusion via partial traces.
\end{itemize}

%\clearpage

%=================================================================
\subsection{Super-scattering based diffusion}
%=================================================================	

Now consider the class of super-scattering operators $\ss_*$ that are entropy increasing

\begin{equation}
S(\ss_*\rho)\geq S(\rho).
\end{equation}
Certainly the decoherence super-scattering operator $\ss_D$ falls in this class; so does $ \ss_{D_s}$.
Ditto for the other super-scattering operators, (based on maximal mixing, partial traces, and the like), considered above: $\ss_M$, $\ss_P$, $\ss_{P_s}$, and $\ss_{\{p,U\}}$. 
In particular such entropy non-decreasing super-scattering operators must leave maximally mixed states maximally mixed:

\begin{equation}
\ss_*\left( {I_N\over N} \right)= {I_N\over N}.
\end{equation}
For any one of these entropy non-decreasing $\ss_*$ operators, we now define 

\begin{equation}
\rho_t = e^{-t} e^{t\$_*} \rho; \qquad \hbox{so that} \qquad\dot \rho_t = \ss_* \rho_t - \rho_t.
\end{equation}
Then for the von Neumann entropy we have

\begin{eqnarray}
\dot S_t &=& -\tr(\dot \rho_t \ln \rho_t) 
\nonumber\\
&=& - \tr( [\ss_*\rho_t]\ln \rho_t) + {\tr(\rho_t\ln \rho_t)}
\nonumber\\
&=&- \tr( [\ss_*\rho_t]\ln [\ss_*\rho_t]) + {\tr(\rho_t\ln \rho_t)} - \{ \tr( [\ss_*\rho_t]\ln \rho_t) -\tr( [\ss_*\rho_t]\ln [\ss_*\rho_t])\} 
\nonumber\\
&=& S(\$_*\rho_t) - S(\rho_t) + \{ \tr( [\$_*\rho_t]\ln [\$_*\rho_t]) - \tr( [\$_*\rho_t]\ln \rho_t) \}.
\end{eqnarray}
But the relative entropy inequality (the Klein inequality) is 

\begin{equation}
\tr( \rho [\ln \rho - \ln\sigma]) \geq 0,
\end{equation}
so automatically

\begin{equation}
 \{ \tr( [\ss_*\rho_t]\ln [\$_*\rho_t]) - \tr( [\ss_*\rho_t]\ln \rho_t) \} \geq 0,
 \end{equation}
 whence
 \begin{equation}
 \dot S_t \geq S(\ss_*\rho_t) - S(\rho_t)  \geq 0.
 \end{equation}
 
 That is, as long as the super-scattering operator $\ss_*$ is guaranteed entropy non-decreasing, then so is the Hilbert-space diffusion process $\rho \to \rho_t =  e^{-t} e^{t\$_*} \rho$.  Since the entropy is always non-decreasing, the only thing that $\rho_t$ can possibly converge on as $t\to\infty$ is the maximally mixed state ${I_N\over N}$ with the maximal entropy $\ln N$. 
 That is, the compound object 
 
\begin{equation}
e^{-t} e^{t\$_*} = \exp(t[\ss_* - I_{N\otimes N}]) = \exp(t \Delta)
\end{equation}
is itself a super-scattering process that can be thought of as a diffusion operator on Hilbert space, corresponding to a ``Laplacian''-like operator $\Delta = \ss_* - I_{N\otimes N}$ which has the property $\tr(\Delta\rho)=0$, (in analogy with the property $\int \Delta\rho(x) \; d^3 x = 0$ holding in the continuum for classical probability distributions).
As usual we can define the hidden information as

\begin{equation}
I_\mathrm{correlations} = S(e^{t\Delta} \rho) - S(\rho) \geq 0,
\end{equation}
which is guaranteed positive and monotonic in the coarse-graining parameter $t$.  The $t$ parameter may or may not represent physical time --- we could view the entire construction as a \emph{gedanken-process} in which case $t$ would just be a tuning parameter giving explicit control of the extent of coarse graining.  The only generic upper bound on the hidden information in this context seems to be

\begin{equation}
I_\mathrm{correlations} = S(e^{t\Delta} \rho) - S(\rho) \leq \ln N - S(\rho).
\end{equation}

%=================================================================
\subsection{Partial trace based diffusion}
%=================================================================
Recall that the partial trace process can be viewed as a \emph{nonlinear} process   $\rho_{AB} \to \mathcal{N} \rho_{AB} = \rho_A \otimes \rho_B$.  The nonlinear operator $\mathcal{N}$ is idempotent, $\mathcal{N}^2=\mathcal{N}$. Now consider

\begin{equation}
\rho \to \rho_t = e^{-t} e^{t\mathcal{N}} \rho =  e^{-t}  \{ \rho + (e^t-1) \mathcal{N}\rho\} = e^{-t}\rho + (1-e^{-t}) \mathcal{N}\rho.
\end{equation}
More explicitly

\begin{equation}
 \rho_t = e^{-t} e^{tN} \rho_{AB}  = e^{-t}\rho_{AB} + (1-e^{-t}) \rho_A\otimes\rho_B = (1-s) \rho_{AB} + s\rho_A\otimes\rho_B.
\end{equation}
Here we have set $s=1-e^{-t} \in [0,1]$ as $t\in[0,\infty)$. That is, a diffusion-like process based on the nonlinear operator $\mathcal{N}$ recovers the process of quantum coarse-graining by tuning the partial traces that we had previously considered, see equation (\ref{E:partial-tuning}).
That is, the compound object 
 
\begin{equation}
e^{-t} e^{t \mathcal{N}} = \exp(t[\mathcal{N} - I_{N\otimes N}]) = e^{-t}I_{N\otimes N} + (1-e^{-t}) \mathcal N,
\end{equation}
while it represents a nonlinear process when acting on density matrices, nevertheless has formal similarities to a diffusion process. 
As usual we can define the hidden information as

\begin{equation}
I_\mathrm{correlations} = S\left(e^{t\{\mathcal{N} - I_{N\otimes N}\}} \rho\right) - S(\rho) \geq 0,
\end{equation}
which is guaranteed positive and monotonic in the coarse-graining parameter $t$.  Again, the $t$ parameter may of may not represent physical time --- we could view the entire construction as a \emph{gedanken-process} in which case $t$ would just be a tuning parameter giving explicit control of the extent of coarse graining. The only generic upper bound on the hidden information in this context seems to be

\begin{equation}
I_\mathrm{correlations} = S\left(e^{t\{\mathcal{N} - I_{N\otimes N}\}} \rho\right) - S(\rho) \leq \ln N - S(\rho).
\end{equation}

%=================================================================
\section{Discussion}\label{S:discussion}
%=================================================================	

We see that we can model the coarse-graining procedure, in a quantifiable and controllable manner, by following different methods depending on what aspect of the system one is taking into account. Classically, when considering the Shannon entropy, we have studied coarse-graining allowing the system change (or not) between continuum and discretium character, and also differentiating situations where we introduce a tuneable smearing parameter to track the change from continuum to discretium. We have studied coarse-graining that involves diffusion, density-density correlations in a box, (depending on a smearing parameter or not), and correlation between the states of two boxes, obtaining in every cases a controllable and monotonic flux of entropy.

On the other hand we have considered the quantum von Neumann entropy, coarse-grained by maximal mixing or by partial traces, tuneable or not, asymmetric in the treatment and also combined with maximal mixing. That corresponds to blurring process depending on a smearing parameter, and choice of separate subsystems that ``decide'' not to look each other. In all cases we have found a positivity of entropy flow and generic upper bound.

We have also studied the coarse-graining induced by decoherence, either partial or complete, allowing us to either remove all off-diagonal terms or control the extent to which we partially suppressed them We have obtained also a consistent definition of the hidden information and an upper bound for it, one that corresponds (as it should) to complete decoherence.

We have introduced an analogy with the Hawking's super-scattering operator that allows us to define a generic class of quantum coarse-grainings. Finally, we have developed a notion of diffusion in Hilbert space,  defined by means of super-scattering or partial traces, again finding a consistent and controlled definition of the entropy flow.

It is important to note that all these models give an explicit and calculable idea of how the entropy flow increases and can be controlled when we coarse-graining a system. These models can be useful tools in order to study some physical processes concerning information issues; such as the well-known information puzzle in black hole physics~\cite{info-puzzle,Hawking:1974,Hawking:1975,Mathur:2008,Mathur:2009,Alonso-Serrano:2015a,Alonso-Serrano:2015b,Alonso-Serrano:2016,Alonso-Serrano:2017},
the ontological status of Bekenstein's horizon entropy~\cite{Bekenstein:1972,Bekenstein:1973,4-laws,Bombelli:1986,Srednicki:1993,Mukohyama:1996}, and the very existence of horizons themselves~\cite{weather,thermality,observability}.

%=================================================================
\vspace{6pt} 
%=================================================================
\acknowledgments{%\textbf{Acknowledgments:} 
AA-S is supported by the grant GACR-14-37086G of the Czech Science Foundation. \\
MV is supported by the Marsden fund, administered by the Royal Society of New Zealand.
}
%=================================================================	

Both authors contributed equally to this article.
%=================================================================

The authors declare no conflict of interest.
%=================================================================	
%\bibliographystyle{mdpi}
%=================================================================
\renewcommand\bibname{References}
%%MDPI internal note: new layout%% redefinition removed
%=================================================================	

%=================================================================	
\end{document}